# Thermodynamic limits of information storage under persistent noise

Katsuaki Tanabe[1,2,3,*]

[1] Department of Chemical Science and Engineering, Kyoto University, Nishikyo, Kyoto 615–8510, Japan

[2] School of Chemical Science and Technology, Kyoto University, Nishikyo, Kyoto 615–8510, Japan

[3] Kyoto MPI Inc., Sakyo, Kyoto 606–8501, Japan

[*] E-mail: tanabe@cheme.kyoto-u.ac.jp

ORCID: 0000-0002-0179-4872

**Abstract**

Writing information and keeping it are physically distinct tasks. When noise cannot be suppressed by the energy landscape, active repair can dissipate free energy throughout storage. How densely can information be retained under a finite repair-speed budget, and at what thermodynamic cost? Here we establish both limits for classical binary memories subject to independent symmetric noise and bounded single-bit repair. The converse permits arbitrary correlations, collective encodings, auxiliary states and time-dependent repair, without access to an external copy of the message. The maximum information density equals the Shannon capacity of a binary symmetric channel whose error probability is fixed by the repair-to-noise ratio. Below capacity, we obtain an explicit minimum maintenance dissipation rate, which diverges logarithmically as capacity is approached. Autonomous linear-code memories attain the frontier in an effective Markov model. At fixed coding slack they support exponentially long retention; logarithmically sized repair blocks suffice for polynomial storage times at fixed dissipation accuracy. Coupling the retention bound to a stochastic speed limit reveals competing costs of rapid computation and prolonged storage, illustrated by an autonomous parity computation. These findings establish a quantitative link between information-theoretic capacity and the thermodynamic cost of reliable information retention under persistent noise.

A computer must not only transform information but also retain it between operations. Physical systems can exploit spontaneous relaxation or consume free energy to protect information against fluctuations. The thermodynamics of computation has addressed the cost of transforming information, from Landauer's principle and logically reversible computing to stochastic thermodynamic descriptions of finite-time operations[1–10]. Recent studies have demonstrated computation using thermodynamically favoured molecular configurations[11,12] and established general time–cost–error bounds for thermodynamic operations[13]. The energetic cost of a transformation alone does not describe a device that must continually protect its information. In an actively repaired memory, dissipation can accumulate with storage time even when the physical probability distribution is stationary.

Thermodynamic uncertainty relations constrain the precision of nonequilibrium currents[14,15]. Related approaches unify stochastic speed limits with uncertainty relations and distinguish housekeeping from excess dissipation[16–18]. Finite-time information processing also exhibits thermodynamic trade-offs between measurement and feedback[19]. The present work instead asks how much information a continuously repaired memory can retain, and at what sustained dissipation rate.

Stochastic thermodynamics provides minimum costs for maintaining prescribed nonequilibrium distributions[20,21], including approaches to electronic and mechanical memories[22,23]. Other work has bounded the initial nonequilibrium resources required for accurate information processing[24], the thermodynamics of molecular copying errors[25] and the thermodynamic costs of communication and coding[26]. A prescribed physical distribution and a prescribed amount of decodable information are different constraints. The latter admits many encodings and physical distributions. Moreover, making repair arbitrarily fast can remove its own entropy-production penalty, whereas a finite repair-rate budget may limit both the achievable information density and the minimum dissipation. Conversely, Shannon's coding theorem[27] and early work on reliable memories made from unreliable components[28] establish information capacities without specifying the thermodynamic cost of finite-rate repair. Lower bounds for noisy circuits and decision trees quantify the logical resources required for reliable computation[29–31], rather than the sustained entropy production of a memory. Recent work on Markovian quantum memories has also established capacity and lifetime bounds[32]. Here we determine the optimal cost over admissible encodings and dynamics for classical memories with unavoidable noise and bounded repair. The capacity converse allows correlated errors and time-dependent repair, and the exact dissipation–density frontier is attained by autonomous coded dynamics. This capacity is a kinetic limit as well as an information-theoretic one; approaching it entails a thermodynamic singularity. Our effective model provides a benchmark for physical implementations, excluding the internal cost of a decoder.

## Physical model and operational capacity

Consider $M$ binary memory variables $S = (S_1, \ldots, S_M)$ and additional physical variables $A$, with full state $Z = (S, A)$ in a finite state space for each $M$. We use the conventional jump-process entropy production for states even under time reversal, and assume nonexplosive dynamics and a well-defined entropy balance. A uniformly distributed $K$-bit message $X$ is encoded in the initial state. The decoder reads $S$ alone, and the initial message label is not an external control signal: all messages share the same generator on $Z$. The initial preparation is not part of the storage interval of duration $\tau$. We require small decoding error throughout that interval; for achievability we impose the stronger requirement that the probability of any decoder change during the interval tends to zero, starting from a correctly decoded state.

Each bit is subject to an independent symmetric noise channel that flips it at rate $\lambda > 0$ in either direction (Fig. 1). This channel cannot be switched off by repair or control. Other transitions changing $S$ flip at most one bit at a time, with the aggregate outgoing repair rate for any given bit bounded by $r$ at each full state. Repair rates may otherwise depend on the full state and time. Auxiliary-only transitions are unrestricted. All physical channels are resolved in the stochastic entropy production, rather than combined into a single effective transition rate. We use dimensionless entropy (physical entropy divided by $k_{\mathrm{B}}$), natural logarithms, and the binary entropy $h_2(q) = h(q)/\ln 2$, where $h(q) = -q\ln q - (1-q)\ln(1-q)$. Supplementary Note 1 distinguishes this model from passive finite-energy memories. The entropy-production convention follows the channel-resolved stochastic thermodynamic formulation[6,33]. Here $t$ is elapsed time, $q$ is a binary-state probability, $k_{\mathrm{B}}$ is the Boltzmann constant, $H$ denotes Shannon entropy, and $I$ denotes mutual information. The index $i$ labels a physical bit, and $S_i$ denotes its value.

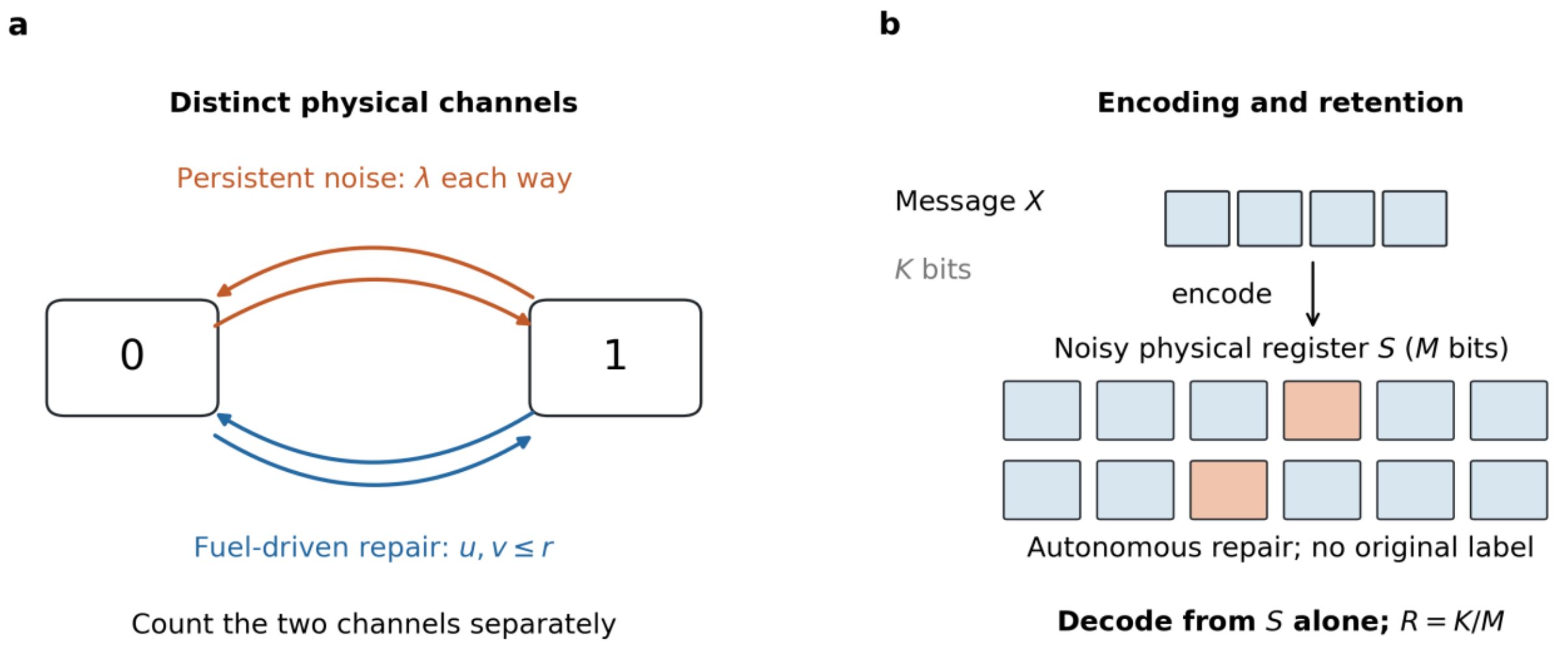

**Fig. 1 | Physical model of an actively stabilized memory. a,** Symmetric noise and controlled repair act through distinct channels on the same bit. In the general model both repair rates are state dependent and bounded above by $r$. **b,** A $K$-bit message is encoded in $M$ noisy physical bits. The repair generator depends on the physical state but not on a separately supplied message label. The decoder need not assign a unique error-free physical configuration to each message. The rates $u$ and $v$ in panel a denote the two repair directions.

The kinetic resource is the dimensionless ratio $\rho = r/\lambda$. Define $q_* = (2+\rho)^{-1}$ and the information density $R = K/M$. Our first result is the asymptotic capacity

$$R_* = 1 - h_2(q_*). \qquad (1)$$

For fixed $\lambda, r > 0$, consider sequences with $M \to \infty$, $K/M \to R$, $\lambda\tau \to \infty$ and decoding error tending to zero uniformly over the storage interval. Equation (1) is an upper bound on $R$ for all such sequences. Every $R < R_*$ is achievable in the model class, including for storage intervals growing as any fixed power of $M$. The capacity is a supremum of achievable densities, not a claim of permanent storage at finite $M$ or of finite dissipation at $R_*$.

The appearance of a binary symmetric channel (BSC) capacity can be understood from an isolated bit whose erroneous value is repaired at the maximum possible rate. With no reverse repair, its stationary error probability is $q_*$. A key point is that the converse does not assume independent bits or a preassigned correct physical configuration. The converse follows from an entropy inequality for arbitrary conditional distributions. Specifically, let

$$D(t) = \ln 2 - \frac{H(S_t \mid X)}{M}, \qquad D_* = R_* \ln 2.$$

The entropy balance yields the comparison inequality

$$D(t) \le D_* + (D(0) - D_*)e^{-(2\lambda+r)t}. \qquad (2)$$

Because $I(X;S_t) \le MD(t)$, and a reliable decoder implies $I(X;S_t) \simeq K\ln 2$, equation (2) gives the capacity converse (Methods). For a finite message error probability $\varepsilon$, the corresponding exact constraint is obtained using Fano's inequality[34,35], $I(X;S_t) \ge K\ln 2 - h(\varepsilon) - \varepsilon\ln(2^K - 1)$, for $\varepsilon \le 1 - 2^{-K}$. No strong converse at nonvanishing block error is asserted. Supplementary Note 2 and Supplementary Fig. 1 give corresponding bounds for exact physical readout and for disjoint one-bit decoding blocks.

# Exact dissipation–density frontier

Let $\Sigma_Z$ denote integrated physical entropy production and the overbar denote a time average. The subscript opt denotes the optimum over admissible memories. All entropy-production rates are expressed in dimensionless entropy per unit time. For $0 < R < R_*$, let $q_R \in (q_*, 1/2)$ solve $h_2(q_R) = 1 - R$. We find the asymptotically optimal rate of physical entropy production per memory bit:

$$\frac{\overline{\sigma}_{\text{opt}}}{\lambda M} = \Psi_\rho(R), \qquad (3)$$

$$\Psi_\rho(R) = (1 - 2q_R)\ln\frac{\rho(1-q_R)}{(\rho+2)q_R - 1}. \qquad (4)$$

Here the optimum means the infimum of limiting normalized time-averaged rates $\Sigma_Z/(\lambda M\tau)$ among the high-reliability sequences described above. The boundary correction for the initial and final state is included in the converse. Initialization, encoding, the final readout and the internal implementation of an ideal nonlocal repair rule are not included in the constructive upper bound. Equations (3) and (4) therefore give the exact optimum for the stated effective Markov model, not a universal cost of all physical memories. Figure 2 summarizes the capacity and frontier.

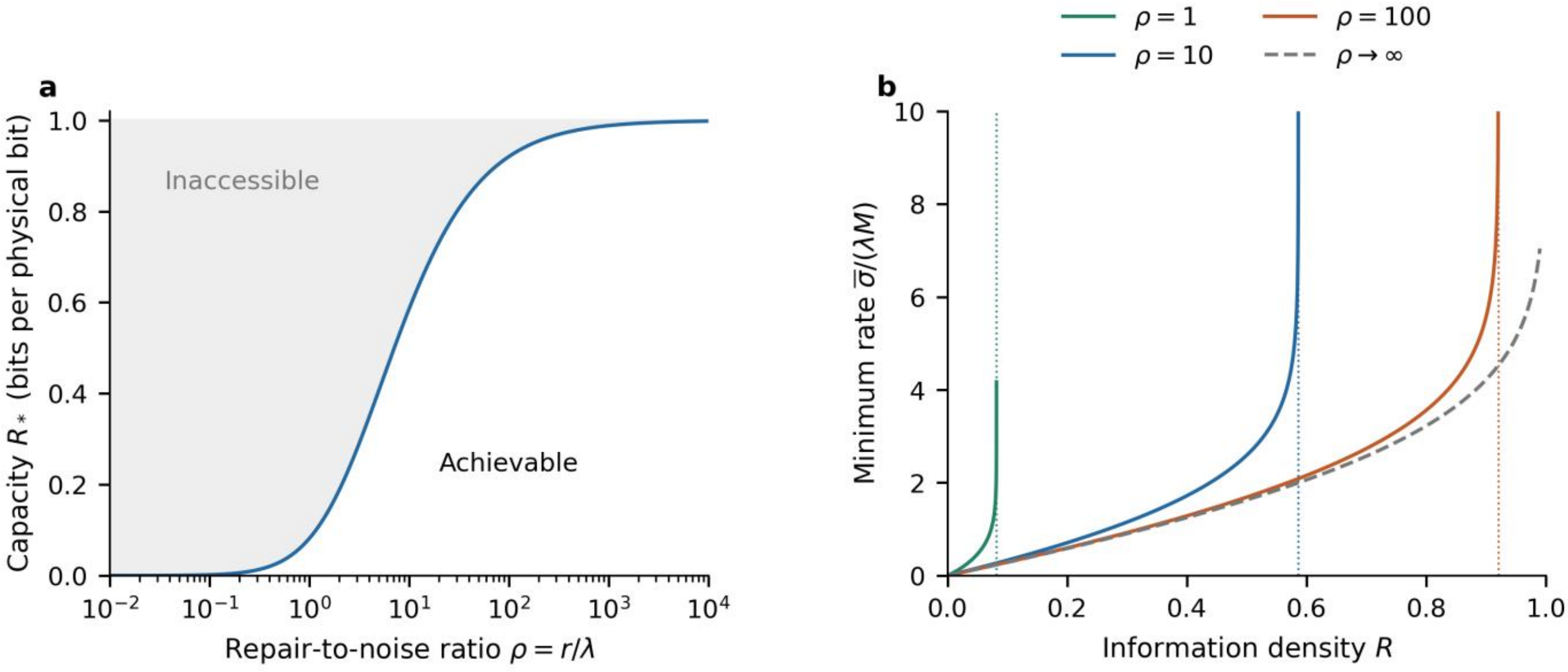

**Fig. 2 | Capacity and minimum maintenance dissipation. a,** Equation (1) gives the maximum asymptotic density of reliably retained information. The shaded region lies above the capacity. **b,** Exact normalized minimum entropy production rate, equation (4), for three finite repair-to-noise ratios. Vertical dotted lines mark the corresponding capacities. The dashed curve is the limit of unconstrained repair speed. All curves are analytic, not fitted to simulations.

The proof uses an entropy balance for the conditional distribution of a bit given the other bits and the message. If its less probable value has probability $q$, symmetric noise generates conditional entropy at rate $\lambda(1-2q)\ln\big((1-q)/q\big)$. Repair must counteract this increase while respecting the finite rate budget. Its own entropy production cannot be neglected when the repair flow approaches its maximal value. We minimize that production using a flux inequality, apply a convex supporting function in the conditional entropy deficit, and average over configurations and time (Methods). The proof allows correlated errors and arbitrary state-dependent repair, rather than assuming a binary symmetric channel a priori. The achievability construction below recovers precisely that channel in the large-memory limit.

For infinite repair speed at fixed $R$, equation (4) reduces to $(1-2q_R)\ln\big((1-q_R)/q_R\big)$. This is the binary nonlinear entropy-production function related to the improved log-Sobolev inequality on the hypercube[36]. The finite-rate result contains an additional kinetic constraint. In the dilute-information limit $\Psi_\rho(R)/R \to 4\ln2(1+1/\rho)$, whereas at the upper capacity boundary

$$\Psi_\rho(R) = \frac{\rho}{\rho+2}\ln\frac{1}{R_* - R} + O(1). \qquad (5)$$

Thus the approach to capacity incurs a logarithmically diverging maintenance cost, even though the information density itself is finite. The coefficient is determined by the nonzero net noise current that repair must compensate at maximal speed. A finite reverse repair current is essential for finite physical entropy production; approaching the capacity forces that current to zero. For $\rho = 10$, $R_* = 0.58618$ and $\Psi_{10}(0.5) = 2.59285$. At the same density, $\Psi_{100}(0.5) = 1.68778$. These values illustrate that a faster repair mechanism can reduce dissipation at a fixed density, while also enabling larger densities. Unlike the target-distribution maintenance bounds[20,21], the frontier optimizes over encodings while explicitly constraining the repair rate.

The bound extends to finite intervals and time-dependent retention. For any chosen $0 < R_0 < R_*$ such that $\tau^{-1}\int_0^\tau I(X;S_t)\,dt \ge MR_0\ln2$, we prove

$$\Sigma_Z \ge \lambda M\tau\Psi_\rho(R_0) - \theta_{R_0}\Delta H(S \mid X) + \Delta I(X;Z). \qquad (6)$$

The positive dimensionless coefficient $\theta_{R_0}$ is given in Methods. The final two terms distinguish dissipation of the physical ensemble from that of an ensemble conditioned on the initial message. The two terms are bounded in magnitude by $\theta_{R_0}M\ln2 + K\ln2$ and are negligible in the long-time normalized rate at fixed $R_0$. Equation (6) does not require a stationary distribution, nor does it require a constant information density at each instant. The inequality relates the time spent storing information to the integrated physical cost of retention. Supplementary Note 3 provides additional finite-time bounds and supporting-inequality checks. Here $\Delta$ denotes final minus initial, and $R_0$ is the prescribed lower bound on the time-averaged information density.

# Attainability with coded nonequilibrium dynamics

We construct a family of autonomous effective memories based on binary linear codes. A realization of the binary register is denoted by $s$. Let $\mathcal{C}$ be a linear code of dimension $K$ and length $M$, and define $d_{\mathcal{C}}(s)$ as the Hamming distance of $s$ to the nearest codeword. Keep the symmetric noise at rate $\lambda$ and choose repair rates $r$ for flips decreasing $d_{\mathcal{C}}$ and $b$ for flips increasing it or leaving it unchanged, where $0 < b < r$. No transition refers to the original message. Fix

$$q = \frac{\lambda+b}{2\lambda+r+b}, \qquad q_* < q < q_R. \qquad (7)$$

The aggregate state dynamics satisfies detailed balance with the distribution

$$\pi(s) = \mathcal{Z}^{-1}\left(\frac{q}{1-q}\right)^{d_{\mathcal{C}}(s)}. \qquad (8)$$

Nevertheless, noise and repair carry opposite stationary currents, producing entropy continuously. Let $T$ denote the temperature and $\Delta\mu$ the chemical potential difference between fuel and waste. In an isothermal reaction-network realization with equal bare configuration energies, a repair transition decreasing the distance to the code can be coupled to a fuel reaction with chemical affinity $\Delta\mu/(k_{\mathrm{B}}T) = \ln(r/b)$; equal-distance flips can be driven reversibly. A separate symmetric noise reaction remains present. The construction provides local detailed balance on the configuration graph. Such a realization need not be spatially local and may require a prohibitive number of configuration-dependent reactions. In equation (8), $\pi$ is the stationary distribution, and the denominator is the partition sum over all configurations. The distinction between aggregate detailed balance and channel-resolved currents follows the stochastic thermodynamic framework[6,33,37].

Choose one minimum-weight coset leader for each syndrome and decode by subtracting that leader. The stationary distribution is uniform over decoded messages, and within a given decoding basin it is a Bernoulli error distribution conditioned on its coset leaders. Starting with this distribution conditional on the intended message prepares the encoded memory. The physical distribution averaged over messages is exactly stationary. Its channel-resolved entropy production rate is

$$\sigma_{\mathrm{ss}} = \lambda(A_{+} - A_{-})\ln(r/b), \qquad (9)$$

where $A_{+}$ and $A_{-}$ are the stationary mean numbers of available flips increasing and decreasing the code distance, respectively. The symbol $\sigma_{\mathrm{ss}}$ denotes the stationary entropy production rate.

Let $W$ denote the number of errors in an independent Bernoulli sample of $M$ bits with error probability $q$. For a uniformly random $K$-dimensional binary linear code, the average Bernoulli probability of being outside the zero-codeword basin or on its inner boundary is bounded by

$$\delta_M = \mathbb{E}_{W\sim\mathrm{Bin}(M,q)}\min\left\{1, \frac{(2^K-1)V_M(W+2)}{2^M-1}\right\}, \qquad (10)$$

where $V_M(w) = \sum_{j=0}^{\min(w,M)}\binom{M}{j}$. At least one code has bad mass at most $\delta_M$. If $R < 1 - h_2(q)$, the bound decreases exponentially with $M$. Stationarity and the rate bound then guarantee that the probability of any decoder change during the interval is at most

$$P_{\mathrm{fail}}(\tau) \le \min\left\{1, M(\lambda+r)\tau\frac{\delta_M}{1-\delta_M}\right\} \qquad (11)$$

when $\delta_M < 1$. Hence the storage reliability holds for polynomially growing durations, not just at an isolated observation time. For any fixed design strictly below its binary symmetric channel capacity, it also holds for sufficiently slowly exponentially growing durations (Methods). The exponent depends on the coding slack and need not stay positive when the frontier is approached. For these same codes, $A_{+}/M \to 1-q$ and $A_{-}/M \to q$; the normalized rate in equation (9) converges to $(1-2q)\ln(r/b)$. Taking $q \uparrow q_R$ after the large-memory limit, or by a sufficiently slow diagonal sequence, yields equations (3) and (4). No extra knowledge of the stored message is used to attain the bound. The syndrome-space evaluation of this construction is detailed in Supplementary Note 4.

Global decoding across the entire memory is not required to approach the frontier. At any fixed $R < R_{*}$ and any fixed positive excess dissipation per bit, the memory can be partitioned into independently repaired blocks of length $m = O(\ln M)$ for storage times $\lambda\tau$ polynomial in $M$ (Methods). The scaling follows by applying equation (11) to each block and a union bound to the whole message. The prefactor in the block length can be large near capacity. The construction reduces the scope of the ideal repair rule from the entire memory to a growing block, without establishing a fixed-range, finite-degree physical decoder. Supplementary Note 5 quantifies the block-size and dissipation guarantees. Low-density parity-check and expander codes provide algorithmic precedents for sparse decoding[38,39]; the present block construction does not establish their physical maintenance cost.

## Reliability over an entire storage interval

We tested the stationary construction using explicit linear codes of length 16, 24, 32 and 40, with $R = 0.5$, and solved the dynamics on their syndrome spaces. At $q = 0.02$ and $\rho = 100$, the instantaneous BSC decoding success probability increases from 98.84% at $M = 16$ to 99.92% at $M = 40$. Yet the corresponding probability of retaining the same decoded message throughout $\lambda\tau = 1$ is only 5.64% and 69.46%, respectively (Fig. 3a). Continuous operation is thus a more stringent requirement than one-time decoding. The normalized stationary entropy production rates at $M = 16,24,32,40$ are 4.3390, 4.2996, 4.3564 and 4.3590, close to the independent-bit value of 4.3639 at the chosen $q$; they are not estimates of the minimum at $R = 0.5$. Further numerical conditions and results are given in Supplementary Table 1. Extended Data Fig. 1 compares the stationary rates and whole-interval failure probabilities across additional design parameters.

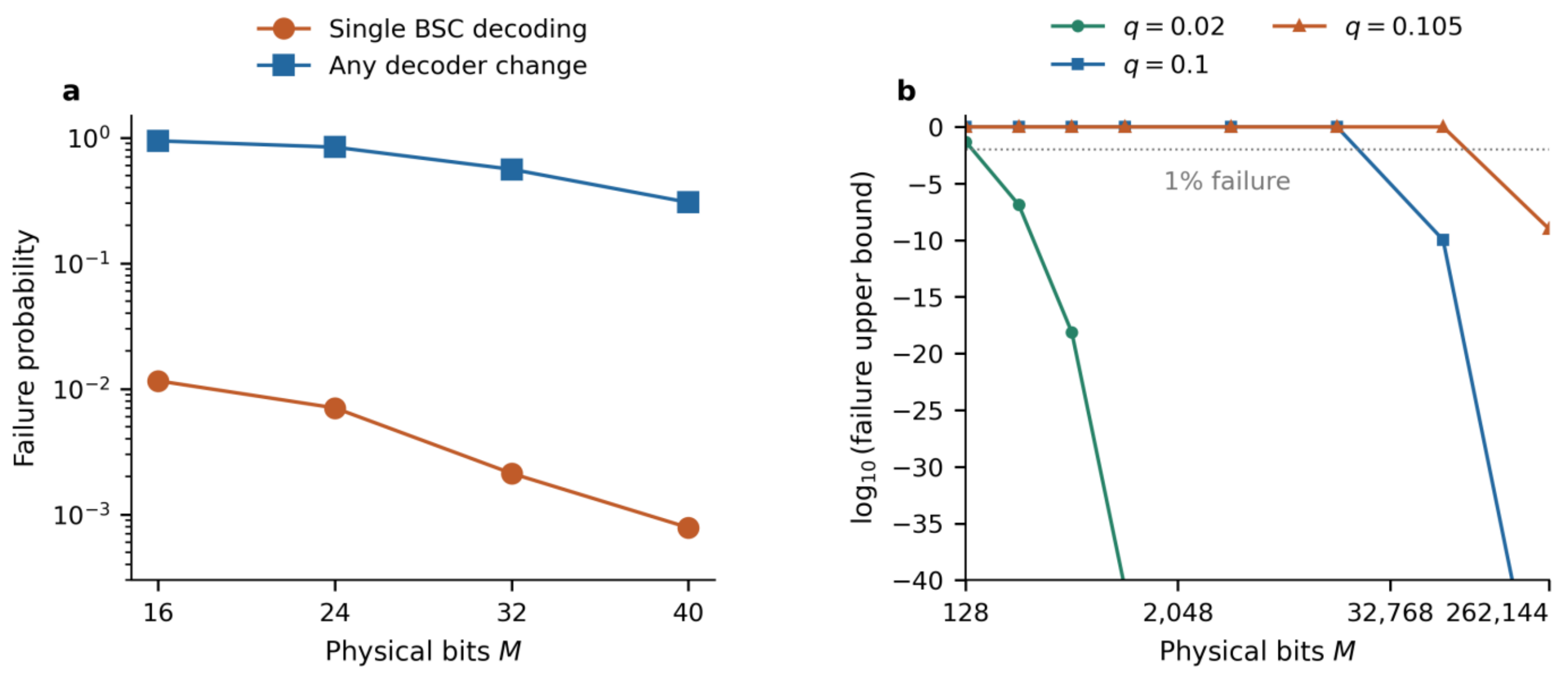

**Fig. 3 | Instantaneous decoding and continuous retention are distinct. a,** For randomly generated systematic linear codes with $K = M/2$, $\rho = 100$ and design error parameter $q = 0.02$, the single BSC decoding error is compared with the probability of at least one decoder change in a stationary memory over $\lambda\tau = 1$. The two smaller retention probabilities are computed from killed master equations; the two larger ones use 5,000 independent uniformization trajectories (error bars: 95% Wilson binomial intervals). The code is fixed for each size; these are not code-ensemble averages. **b,** Existence bounds from equations (10) and (11) for $K = M/2$, $\rho = 100$, and $\lambda\tau = 1$. These are analytic finite-size guarantees for the existence of a code, not results of simulations of individual large codes. Values below the vertical plotting range are not shown.

For larger sizes, the existence guarantee avoids exhaustive simulation (Fig. 3b). For example, at $M = 512$, $R = 0.5$, $\rho = 100$, $q = 0.02$ and $\lambda\tau = 1$, the failure bound is $7.28 \times 10^{-19}$. A closer approach to capacity requires substantially larger codes. At $q = 0.105$, $M = 262{,}144$ and the same density and rate ratio, the bound is $9.71 \times 10^{-10}$, while the asymptotic normalized dissipation of the construction is 1.75466, just 3.96% above the frontier. For such large sizes the values are existence and asymptotic rate estimates, not measurements of a realized memory. The distinction matters because implementing a decoder may itself require considerable physical resources.

As an example of the block construction, take $M = 2^{20}$, $R = 0.5$, $\rho = 100$ and $\lambda\tau = M$. With $q = 0.08$, 128 independent blocks of 8,192 bits suffice for a whole-message failure bound of $1.07 \times 10^{-7}$; their normalized stationary dissipation differs negligibly from 2.14475 (27.1% above the frontier). With $q = 0.10$, eight blocks of 131,072 bits give a failure bound of $4.48 \times 10^{-19}$ and a normalized rate of 1.82448 (8.10% above the frontier). The estimates are existence guarantees, not implemented decoders. The rapid growth of the required block size illustrates the engineering distinction between attaining an information-theoretic limit and implementing its repair rules. The full block-size comparison is given in Supplementary Table 2. Extended Data Fig. 2 visualizes the block-size versus excess-dissipation trade-off for the same finite-size guarantees. Earlier work establishes reliable storage from unreliable components[28] and quantifies the cost of communication and coding[26]; the present estimates quantify sustained repair and the probability of a decoder change within a storage interval.

## Consequences for finite-time computation

Information retention competes with the kinetic cost of computation. As a restricted example, consider a sequential computing architecture with a progress coordinate whose initial and successful final locations are separated by at least $L$ graph steps. Its transitions are physically distinct from the noise and repair channels and do not directly change the noisy register $S$. Let $r_c$ bound the total progress-changing rate, and let $D = L(1 - \varepsilon_c)$, where $\varepsilon_c$ is the probability of failing to finish by the deadline. A graph Wasserstein speed limit[40] bounds the progress-channel entropy production by $2D\operatorname{artanh}\big(D/(r_c\tau)\big)$. If the memory satisfies the retention criterion of equation (6), we obtain

$$\Sigma_Z \geq 2D\operatorname{artanh}\left(\frac{D}{r_c\tau}\right) + B\tau - C_{\mathrm{bd}}, \qquad (12)$$

where $B = \lambda M \Psi_\rho(R_0)$ and $C_{\mathrm{bd}} = \theta_{R_0}\Delta H(S \mid X) - \Delta I(X;Z)$. The two channel contributions are additive under the stated separation. Replacing $C_{\mathrm{bd}}$ by its duration-independent upper bound, the right-hand side has a unique minimizing duration for $B > 0$:

$$\tau_* = \frac{D}{r_c}\sqrt{1 + \frac{2r_c}{B}}. \qquad (13)$$

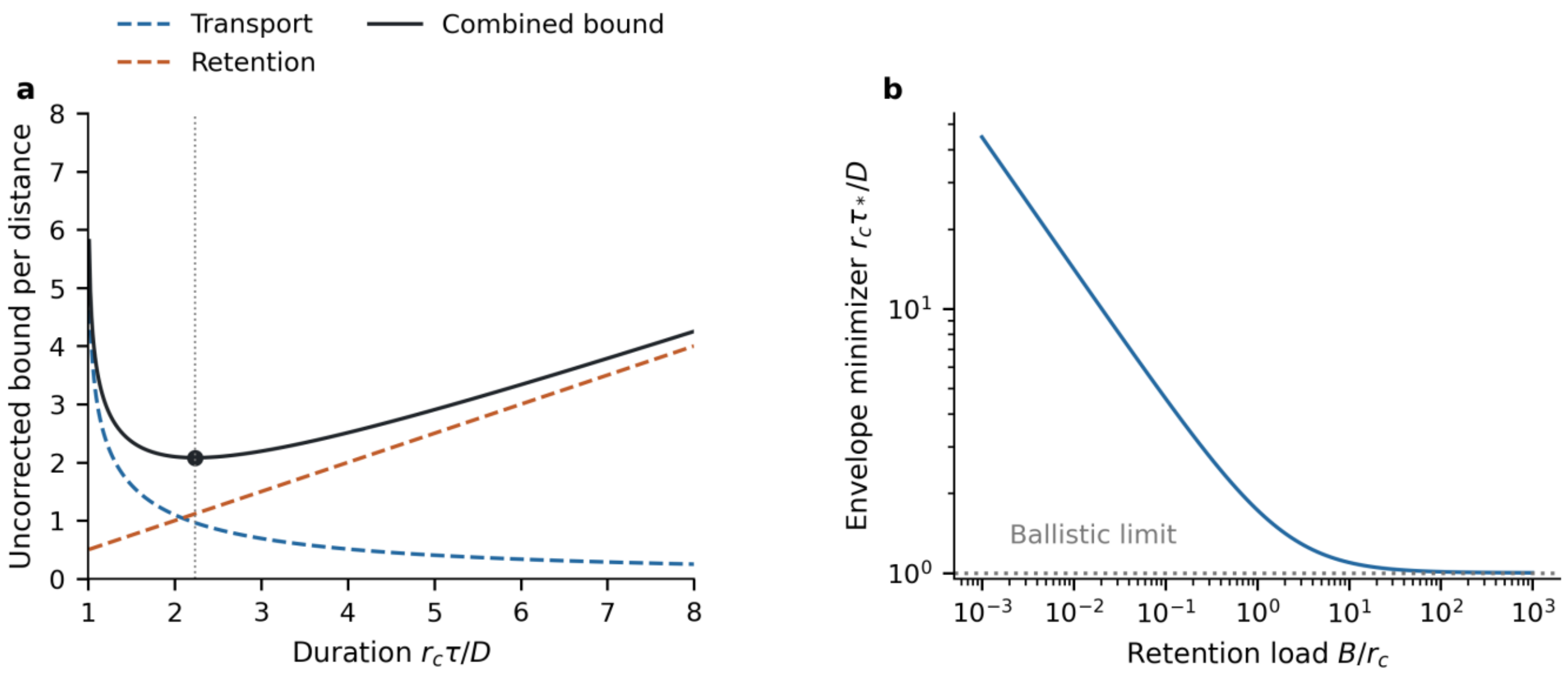

**Fig. 4 | Retention competes with the speed cost of computation. a,** Dimensionless contributions to equation (12), excluding the boundary correction, for $B/r_c = 0.5$. The marked minimum lies above the ballistic time $D/r_c$. **b,** Minimizer of the lower envelope as a function of the retention load. This is a bound, not a demonstrated optimum of a complete computer. The additive relation assumes that progress transitions do not change the noisy register.

Slowing computation reduces the transport term but increases the cost of maintaining information (Fig. 4). Approaching the storage capacity at fixed kinetic resources forces the minimizing duration of this lower envelope toward the ballistic limit, $D/r_c$. For sequential architectures with $L$ and $M$ both proportional to problem size and nonzero fixed density and noise, the memory term alone gives a quadratic lower bound in problem size, provided the information is retained throughout the computation. The quadratic bound is not a universal bound for all algorithms: an architecture that discards unnecessary information, retains it passively, or performs genuinely parallel operations may have a different information–time volume. The bound cannot be added to a speed limit for the same memory-changing transitions without an additional argument. Supplementary Note 6 discusses the passive-computation comparison and the cost of retaining unnecessary information. The trade-off is complementary to finite-time parallelization[10] and reversible checkpoint scheduling[41], which concern the execution and cleanup of logical operations rather than persistent maintenance noise. The role of activity in speed limits for classical stochastic processes was established in earlier work[42]. A complementary analysis of Gibbs free-energy speed-limit tightness in deterministic chemical reaction networks identifies stoichiometric projection and kinetic-mode heterogeneity as distinct sources of nonsaturation[43]. In the present stochastic computing model, the minimizer of the additive lower envelope likewise must be distinguished from an attained optimum of the full device.

The transport contribution is attainable to leading order in a restricted computing model. Store $L$ input bits in the autonomous coded memory above. A biased nearest-neighbour clock reads one decoded input bit at each forward step and applies the exclusive-OR operation to a separate one-bit accumulator; a backward step undoes the same operation. On the event that the decoded input never changes, the accumulator at completion contains its parity. The dimensionless clock drift bias is $\delta$. For any fixed $0 < \delta < 1$, a deadline $\tau = L/(r_c\delta)$ can be met with error tending to zero as $L \to \infty$. The combined entropy production is $\sigma_{\mathrm{ss}}\tau + 2L\operatorname{artanh}\delta + o(L)$ (Methods). Thus the transport term is sharp to leading order on top of the actual stationary memory cost. The accumulator and gate rules are idealized, and this construction neither establishes simultaneous finite-size saturation of equation (12) nor includes the cost of a fault-tolerant output register. Supplementary Table 3 and Supplementary Fig. 2 report finite-clock numerical checks. Studies of Brownian and autonomous clocks[44–46] relate dissipation to timing precision; here the clock is used as a progress coordinate implementing reversible logic.

## Physical scope and implications

The assumptions are essential. Passive memories can store information in long-lived metastable sectors of an equilibrium dynamics, as illustrated by self-correcting classical and quantum codes[47]. In such systems the physical energy landscape can suppress the offending transitions themselves. Similarly, a bistable memory refreshed by a controlled protocol[22] is not necessarily a memory with an unavoidable independent symmetric error channel. Our bound concerns the distinct class in which that channel remains operative independently of repair. The bound does not imply that passive equilibrium memories or all thermodynamically favoured molecular computers dissipate at the rates derived here. Conversely, a reversible aggregate generator does not eliminate the positive dissipation of physically distinct noise and fuel-driven repair channels.

The results distinguish three resources: the number of physical bits, the maximum rate of repair, and entropy production. The frontier is convex and vanishes at zero density, so its dissipation rate per logical bit is nondecreasing with density. Increasing redundancy reduces this rate toward the nonzero dilute-limit value stated above; even arbitrarily large redundancy cannot eliminate maintenance dissipation at fixed persistent noise. Increasing repair speed raises the maximum density and decreases the finite-rate penalty. At the density boundary, irreversible repair is approached even though the noise dynamics itself is time-reversal symmetric. The mechanism differs from the high-accuracy divergence for individual time-symmetric logical transformations[48]: here the divergence concerns a storage capacity approached at a fixed repair-speed budget. The treatment of correlated errors, physical rate bounds and collective decoding connects a Shannon-type storage limit to an explicitly thermodynamic optimum. Spatially local realizations, accounting for the complexity and dissipation of a physical decoder and noise processes that are not symmetric independent flips, are natural settings in which to investigate stricter frontiers.

# Methods

## Model conventions and entropy balance

Let $z$ and $z'$ denote configurations, $p$ their probability distribution, and $w$ a channel transition rate. For a channel-resolved continuous-time Markov process, write the entropy production rate[6,33] as the sum over unordered state pairs and channels of $(F-B)\ln(F/B)$, where $F = p(z)w_{zz'}$ and $B = p(z')w_{z'z}$. A channel with nonzero one-way flux and zero reverse flux has infinite instantaneous entropy production. Rates and states may be time dependent through the prescribed generator, but the generator on $Z$ must be common to all $X$. Conditional entropy production $\Sigma_{Z|X}$ means the average over initial message labels of the physical expression with the conditional distributions. Since environmental entropy changes are linear in the initial distribution and $H(Z) = H(Z \mid X) + I(X;Z)$,

$$\Sigma_Z = \Sigma_{Z|X} + I(X;Z_\tau) - I(X;Z_0). \qquad (14)$$

All rates changing a particular bit through repair are bounded in aggregate by $r$. Coarse-graining the auxiliary variables and merging repair channels by the log-sum inequality can only reduce conditional entropy production. The resulting bit-flip fluxes still obey the rate cap even if the effective marginal rates depend on time and on $X$. The coarse-grained entropy balance is exact. Noise remains a symmetric flip at rate $\lambda$ in every conditional ensemble. For each bit $i$, condition on $X$ and the other bits $S_{-i}$, with weight $m = p(X,S_{-i})$. Write $q \le 1/2$ for the smaller conditional probability, $x = \ln\big((1-q)/q\big)$, and let $F,B$ now denote the repair fluxes per unit $m\lambda$ from the less probable value to the more probable value and back. The fluxes obey $F \le \rho q$, $B \le \rho(1-q)$. With $j = F - B$ and $n = 1-2q$, the contribution to $(\lambda M)^{-1}dH(S \mid X)/dt$ is $(n-j)x$, averaged over all $i, X, S_{-i}$ with normalized weights $m/M$. The corresponding entropy production contribution is at least $nx + j\ln(F/B)$. Ties $q = 1/2$ and zero probabilities are treated by continuity (or the extended-real flux expression). Auxiliary-only transitions do not change the marginal conditional entropy of $S$. The conditional-to-marginal entropy balance is related to the initial-distribution dependence of entropy production[49]. The coarse-graining step uses the log-sum inequality[35,37].

## Convex functions for the converse

Define $d = \ln 2 - h(q)$ and $\phi(d) = nx$. The change of variables $q = (1+e^x)^{-1}$ gives $d'(x) = xq(1-q)$ and

$$\frac{d\phi}{dd} = 2 + \frac{2\sinh x}{x},$$

so $\phi$ is increasing and convex. We also require a concavity fact: if $G$ is differentiable, convex on $y \ge 0$, with $G(0) = 0$, $G' \ge 0$ and $yG''(y) \le G'(y)$, then $qG(\theta x)$ is concave in $d$ for any $\theta \ge 0$. Indeed,

$$\frac{d}{dd}\big(qG(\theta x)\big) = -\frac{G(\theta x)}{x} + \frac{\theta G'(\theta x)}{x(1-q)}.$$

Differentiating the right-hand side in $x$ and multiplying by $x^2$ gives the sum

$$G - yG' + \frac{\theta(yG'' - G')}{1-q} - \frac{yG'q}{1-q}, \qquad y = \theta x,$$

which is nonpositive by the hypotheses. The endpoint at $x = 0$ is understood as a limit. In particular $qx$ is concave in $d$ (take $G(y) = y, \theta = 1$).

## Capacity converse

Since $j \le \rho q$, the conditional entropy increase per unit normalized weight is bounded below by $f(d) = \phi(d) - \rho q x$. The preceding convexity and concavity establish convexity of $f$. At $q = q_*$, $f(D_*) = 0$ and $f'(D_*) = 2 + \rho$. Thus $f(d) \ge (2+\rho)(d - D_*)$. The conditional entropy chain rule gives

$$\frac{1}{M}\sum_i H(S_i \mid S_{-i}, X) \le \frac{H(S \mid X)}{M},$$

so the mean deficit $\langle d \rangle$ is at least $D(t)$, which itself is at least $I(X; S_t)/M$. Averaging the supporting inequality yields $\dot{D} \le -(2\lambda + r)(D - D_*)$, proving equation (2). The entropy argument is independent of thermodynamic irreversibility and therefore even allows unidirectional repair. Fano's inequality then gives equation (1) as a necessary asymptotic condition. The entropy chain rule and decoding-error inequality are standard information-theoretic relations[34,35].

## Sharp finite-rate dissipation converse

For $j \ge 0$, minimizing the repair entropy production at fixed net flux uses the maximal forward flux $\rho q$, giving

$$j\ln(F/B) \ge \rho q\, g\left(\frac{j}{\rho q}\right), \qquad g(k) = -k\ln(1-k), \quad 0 \le k < 1.$$

Let $G(y)$ be the maximum of $yk - g(k)$ over $0 \le k < 1$. For $y \ge 0$ the optimizer satisfies $g'(k) = y$; hence $G' = k$, $G'' = 1/g''(k)$ for $y > 0$. The derivative $g''(k) = (2-k)/(1-k)^2$ increases on the domain and $g'(0) = 0$, so $g'(k) \le k g''(k)$ and $yG''(y) \le G'(y)$. The hypotheses of the preceding concavity result hold. For $j < 0$, nonnegativity of repair entropy production and $\theta j x \le 0$ give the same following inequality. For every $\theta \ge 0$,

$$nx + j\ln(F/B) + \theta(n - j)x \ge L_\theta(d), \qquad (15)$$

$$L_\theta(d) = (1+\theta)\phi(d) - \rho q G(\theta x).$$

The right-hand side is convex. Choose any $0 < R_0 < R_*$ and set $d_0 = R_0 \ln 2$, $q_0 = q_{R_0}$, $x_0 = \ln\big((1-q_0)/q_0\big)$, $k_0 = (1 - 2q_0)/(\rho q_0)$ and

$$\theta_{R_0} = \frac{g'(k_0)}{x_0}. \qquad (16)$$

At $d = d_0$, the bound (15) is attained by the formal fluxes $F = \rho q_0$ and $j = n$, with $B = F - j > 0$. Consequently $L_\theta(d_0) = \Psi_\rho(R_0)$. To see that the supporting slope is positive, consider the balanced-current cost $C(d) = \phi(d) + \rho q g\big(n/(\rho q)\big)$ on $0 < d < D_*$. For any fixed $\theta$, $L_\theta(d) \le C(d)$, with equality at $d_0$ for the chosen $\theta$. Their first derivatives coincide there. $C(d)$ equals equation (4) at $R = d/\ln 2$, and is strictly increasing because both its positive factors increase as $q$ decreases. Thus $L_\theta{}'(d_0) > 0$.

Average (15) over the bit and conditional configurations, and integrate over time. If the mean mutual information per bit is at least $d_0$, the mean $d$ over time and configurations is also at least $d_0$. The supporting line for the convex $L_\theta$ then gives

$$\Sigma_{Z|X} + \theta \Delta H(S \mid X) \ge \lambda M \tau \Psi_\rho(R_0).$$

Using (14) gives equation (6). For any sequence with $K/M \to R$, uniformly vanishing decoding error and $\lambda\tau \to \infty$, fix $R_0 < R$ and use Fano's inequality, followed by the long-time limit and $R_0 \uparrow R$. The argument proves the lower bound for equation (3), without interchanging a divergent $\theta_{R_0}$ with the long-time limit. At $R = R_*$ the same procedure for arbitrary fixed $R_0 < R_*$ establishes divergence of the normalized rate. For equation (5), expand $R_* - R = \big(h(q_R) - h(q_*)\big)/\ln 2 = x_*(q_R - q_*)/\ln 2 + O((q_R - q_*)^2)$, where $x_* = \ln\big((1 - q_*)/q_*\big) > 0$. Expanding equation (4) then gives the coefficient $1 - 2q_*$. The dilute limit follows from $d = x^2/8 + O(x^4)$ and expanding the two factors in (4). The same supporting-line argument shows convexity of the balanced-current cost on its feasible interval: each interior point has a global supporting tangent, since the convex supporting function lies below that cost and touches it there. The continuous extension to zero density vanishes. Dividing by the density therefore yields a nondecreasing rate per logical bit.

## Code construction and finite-size bounds

Take a full-rank binary parity-check matrix of size $(M - K) \times M$. The syndrome is the binary parity-check vector of a configuration. For every syndrome $y$ choose one minimum-weight leader $e(y)$, breaking ties consistently; the decoder is $s \oplus e(y)$. The basins are translates of the zero-codeword basin $A_0 = \{e(y)\}$. Neighboring code distances differ by at most one. The aggregate transition rates are $\lambda + r$ down and $\lambda + b$ up, with ratio $(\lambda + b)/(\lambda + r) = q/(1 - q)$; equal-distance edges have equal rates. Hence (8) is an exact stationary distribution and the aggregate chain is reversible. The chain is irreducible at finite $M$ due to noise. The message-conditioned initial distribution $\mu_0$ restricted to $A_0$ is a Bernoulli($q$) error distribution conditioned on the chosen leaders, translated to the encoded codeword. Averaging over a uniform message gives $\pi$.

For an up edge $s \to s'$, the stationary flux equality is $(\lambda + b)\pi(s) = (\lambda + r)\pi(s')$. The positive noise current $\lambda\big(\pi(s) - \pi(s')\big)$ is balanced by an opposite repair current. The cycle affinity is $\ln(r/b)$; equal-distance edges have zero contribution. Summing yields (9). The calculation uses the channel-resolved physical distribution and is not the entropy production of the reversible aggregate chain.

Let $E$ be the set outside $A_0$ or in $A_0$ with at least one neighbor outside it. If $s \in E$ and $w = \mathrm{wt}(s)$, a nonzero codeword lies within Hamming distance $w + 2$ of $s$: outside the basin a competitor is at least as close as zero; on the inner boundary use the same fact for the neighboring state and the triangle inequality. In a uniformly random $K$-dimensional linear subspace each nonzero vector belongs to the code with probability $(2^K - 1)/(2^M - 1)$. A union bound at fixed $s$, clipped at one and averaged over the Bernoulli errors, proves (10). To establish exponential decay, choose $q < a < 1/2$ with $R < 1 - h_2(a)$. On $W \le aM$, the volume factor is bounded by $\exp\Big(M\big(h(a) + o(1)\big)\Big)$, giving a negative exponential rate. The complementary binomial tail also decays exponentially. The argument proves existence without invoking an unproved assumption on boundary mass. In particular, if the bad mass is bounded by exp(−cM) for some fixed positive c, equation (11) implies vanishing probability of any decoder change for storage durations in units of the inverse noise rate up to exp(c′M), for any fixed 0 < c′ < c. This exponential-lifetime guarantee applies at fixed design slack; it does not establish a positive uniform lifetime exponent along a sequence approaching the exact dissipation frontier. The estimate uses the usual random-coding and union-bound arguments for binary channels[27,35].

In the stationary ensemble, the fraction of time spent at an inner basin boundary is at most $\delta_M/(1 - \delta_M)$ by conditioning. A decoder change must depart from such a boundary, and the total exit rate is bounded by $M(\lambda + r)$. The expected number of decoder changes over $\tau$ therefore satisfies the right-hand side of (11) before clipping; Markov's inequality proves (11). Translation symmetry gives identical success probabilities for each initial decoded message. Away from the bad set, all neighbors decode to the same codeword. Hence the numbers of up and down edges are exactly $M - w$ and $w$, respectively. Conditioning the Bernoulli error distribution on $A_0$ changes the expectation of any function bounded between zero and one by at most $\mu_0^{\mathrm{uncond}}(A_0^c) \le \delta_M$, and the contribution from boundary states to a normalized edge count is at most $\delta_M/(1 - \delta_M)$. In particular the normalized up and down counts differ from $1 - q$ and $q$ by vanishing errors. The limiting construction proves (3) by first choosing fixed $q < q_R$ with $b = \big(rq - \lambda(1 - 2q)\big)/(1 - q) > 0$, taking large $M$ and polynomially growing $\tau$, and then approaching $q_R$. The upper bound is autonomous and the physical ensemble is stationary; the construction does not require a controller that reads the initial label. A family of rates and codes can be selected along a diagonal sequence if desired.

## Finite-size dissipation and blockwise repair

Let the Bernoulli bad-set mass of the chosen code be $\eta \leq \delta_m < 1$, with block length $m$. Off the bad set, $\big(A_+(s) - A_-(s)\big)/m = 1 - 2|s|/m$. Both functions take values in $[-1,1]$. Conditioning the Bernoulli distribution on the zero-codeword basin changes the latter mean by at most $2\delta_m$; the conditional mass of the inner boundary is at most $\delta_m/(1-\delta_m)$. It follows from (9) that

$$\left|\frac{\sigma_{\mathrm{ss}}}{\lambda m} - (1-2q)\ln(r/b)\right| \leq \left(2\delta_m + \frac{2\delta_m}{1-\delta_m}\right)\ln(r/b).$$

For a fixed $R < R_*$ and a desired positive accuracy of the frontier value, choose $q_* < q < q_R$ sufficiently close to $q_R$. The random-code estimate gives $\delta_m \leq e^{-cm}$ for some $c > 0$ and all sufficiently large $m$, with $k = \lfloor Rm \rfloor$. Use identical copies of a code attaining this estimate in $J$ independent blocks, so $M = Jm$ and $K = Jk$. The stationary dissipation is additive, and equation (11) with a union bound gives whole-message failure probability at most $M(\lambda + r)\tau\delta_m/(1-\delta_m)$ (clipped at one). For $\lambda\tau = O(M^p)$ with fixed $p > 0$, taking $m \geq C\ln M$ with $C > (p+1)/c$ makes this bound vanish and gives the stipulated normalized dissipation accuracy. One can choose a sequence of sizes divisible by $m$; rounding the block rate does not change the limit. Exact convergence to the frontier may require a growing $C$ as the capacity gap closes. The explicit configuration-dependent rate table has at most $O(M2^m)$ entries, polynomial in $M$ at fixed $C$. Neither this count nor the block construction accounts for the hardware cost or establishes a fixed-range interaction implementation.

## Numerical methods

Systematic parity-check matrices were formed from identity columns and distinct uniformly sampled nonzero columns with the default random generator in NumPy, using seed $20261125 + M$. Coset leaders and minimum distances were determined by breadth-first search on the syndrome graph. Each leader's syndrome and weight were verified independently. Stationary probabilities, the decoder-change hazard and equation (9) were evaluated by summation over all $2^{M-K}$ syndromes. Detailed-balance residuals in the aggregate process were below $10^{-15}$ in absolute probability-flow units. For $M \leq 24$, the killed generator (removing off-diagonal decoder-changing transitions but retaining their escape rates on the diagonal) gives the probability of no decoder change as the sum of the propagated probability vector. For $M = 32{,}40$, exact uniformization with dominating rate $M(\lambda + r)$ and 5,000 trajectories was used. For the Fig. 3a series the simulation seed is $20261125 + M + 20 + 100$. Initial syndromes were sampled from their stationary distribution. Figures report 95% Wilson binomial intervals; Supplementary Table 1 reports binomial standard errors. Ensemble bounds were evaluated with logarithmic binomial coefficients, cumulative log-sums of Hamming-ball volumes and log-sum-exp over the binomial probabilities. These calculations used NumPy 2.3.5 and SciPy 1.17.0.

## Speed-limit corollary

The progress coordinate is unchanged by the noise and repair channels. Its own channels are also assumed not to change $S$; thus they do not contribute to the conditional entropy balance used in (15). Applying the log-sum inequality to those channels in the conditioned ensemble and then marginalizing the message gives at least the entropy production of the coarse-grained progress process. If the start is deterministic, the graph Wasserstein distance from the initial distribution to the deadline distribution is at least $D = L(1-\varepsilon_c)$. The transport bound[40] is $\Sigma_{\mathrm{progress}} \geq 2D\,\mathrm{artanh}(D/\mathcal{A})$, where $\mathcal{A}$ is the expected number of progress jumps. Since $\mathcal{A} \leq r_c\tau$, this establishes the first term of (12) (requiring $r_c\tau > D$ for a finite bound). Adding the distinct channel bound and using (14) once gives (12). Differentiating the duration-dependent expression with a constant upper bound on $C_{\mathrm{bd}}$ yields (13). These statements concern the lower envelope, not its simultaneous attainability for an arbitrary computing architecture.

## Autonomous parity example

Use the stationary coded memory with $K = L$, $M$ proportional to $L$, and a whole-interval failure probability tending to zero for $\tau = O(L)$. Add a clock $h \in \{-L, \dots, 2L\}$ initialized at zero and an ideal bit $Y$ initialized at zero. The clock is a reflecting birth–death chain with constant interior rates $u = r_c(1 + \delta_L)/2$ and $v = r_c(1 - \delta_L)/2$, where $\delta_L = \delta + L^{-1/3} < 1$ for sufficiently large $L$. The forward and backward rates are $u$ and $v$, respectively. Let $X_i(S)$ denote the decoded value of input bit $i$. On crossing the edge $i - 1 \leftrightarrow i$, for $1 \le i \le L$, in either direction, set $Y \mapsto Y \oplus \hat{X}_i(S)$; outside these edges leave $Y$ unchanged. These are reversible maps at fixed $S$, so each channel has the reverse rate $v$ or $u$, respectively. The memory transitions are unchanged and do not alter $h$ or $Y$. If no decoder change occurs, the accumulator equals the parity of the first $\min\{L, \max\{0, h\}\}$ input bits. In particular the accumulator is correct when $h \ge L$.

To evaluate the deadline $\tau = L/(r_c\delta)$, couple the clock to the unrestricted difference of independent Poisson processes of rates $u, v$. Its mean displacement is $r_c\delta_L\tau = L + L^{2/3}/\delta$ and its variance is $r_c\tau = O(L)$. The probability of ending below $L$ is $O\left(L^{-1/3}\right)$ by Chebyshev's inequality. The centred process is a square-integrable martingale; its maximal deviation exceeds a constant times $L$ with probability $O(L^{-1})$ by the maximal inequality. Since the deterministic drift stays a distance of order $L$ from the reflecting endpoints over this interval, the probability that reflection alters the unrestricted path is $O(L^{-1})$. Thus the computation failure probability is bounded by the memory failure probability plus $O\left(L^{-1/3}\right)$, and $\langle h_\tau \rangle = L + o(L)$ (use Cauchy–Schwarz on the reflection event). The count of net forward clock jumps is exactly $h_\tau$.

The marginal physical distribution of $S$ is stationary and independent of $h$. The medium entropy production of the memory channels is therefore $\sigma_{\mathrm{ss}}\tau$, while that of the clock is $\langle h_\tau \rangle \ln(u/v)$. The total system entropy change is $H(h_\tau) + H(Y_\tau \mid S_\tau, h_\tau)$, lying between zero and $\ln(3L + 1) + \ln 2$. Consequently $\Sigma_Z = \sigma_{\mathrm{ss}}\tau + 2L\,\mathrm{artanh}\delta + o(L)$. The coarse-grained speed bound tends to the same leading clock term because the completion probability tends to one. The comparison holds on top of the actual stationary maintenance cost, which can approach the frontier in normalized rate; the comparison does not guarantee saturation of the finite-size boundary-corrected inequality. The ideal output bit and clock-gate mechanism are additional assumptions. For numerical illustration, the finite reflecting clock was propagated using a sparse tridiagonal generator, with column sums and probability normalization checked. A separate script evaluates the block bounds and the supporting inequalities.

## Data availability

Numerical data are available on request from the corresponding author.

## Competing interests

The author declares no competing interests.

## Additional information

Extended data and supplementary material are available at https://doi.org/xxx.

# Extended Data

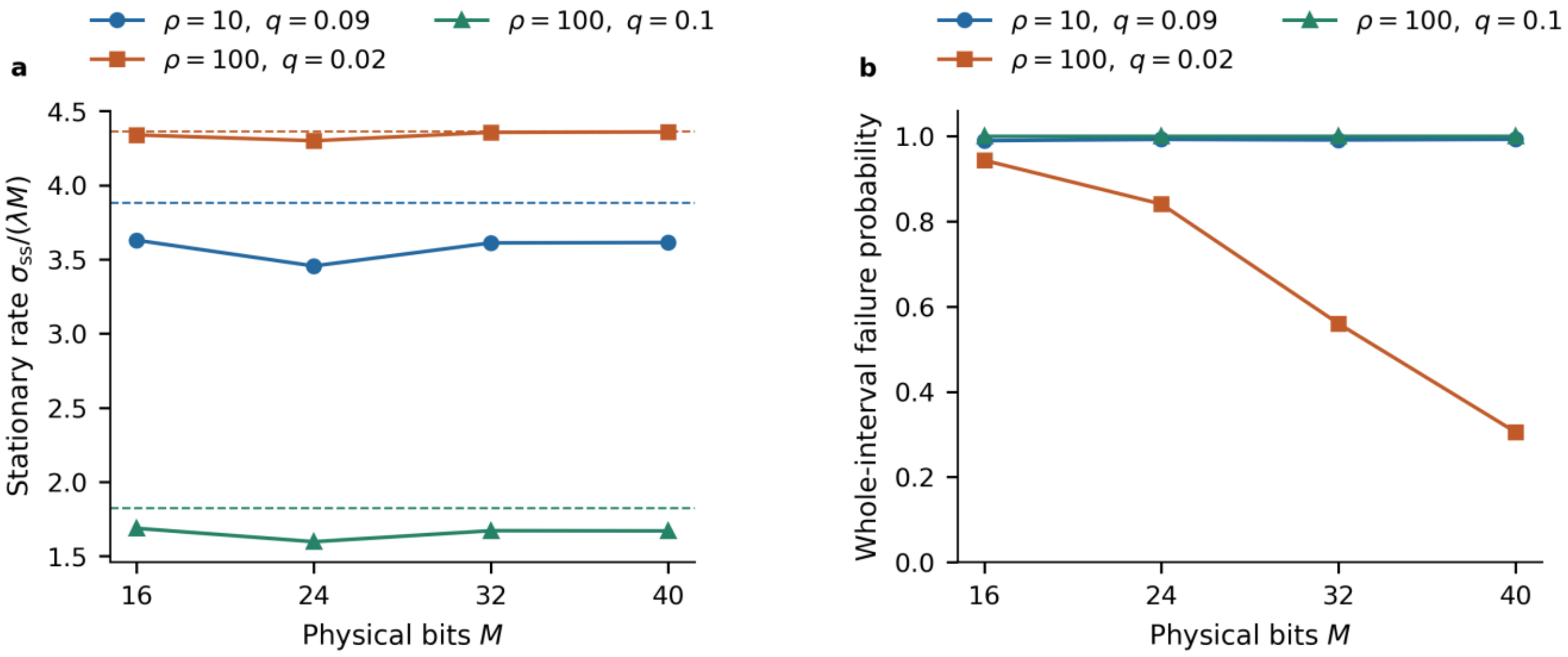

**Extended Data Fig. 1 |** Finite-size dissipation and reliability at additional design parameters. a, Stationary entropy production rate for fixed systematic codes of rate $R = 0.5$; dashed lines are the independent-bit reference rates at the corresponding $q$ and $\rho$. b, Probability of at least one decoder change during $\lambda\tau = 1$ for the same codes. $M = 16$ and 24 use the killed master equation; $M = 32$ and 40 use 5,000 uniformization trajectories, with 95% Wilson binomial intervals (too small to resolve for some points). Values rounded to one in exact calculations or sampled as one do not imply certain failure. These are individual fixed code realizations. The finite-code rate can be lower than the asymptotic optimum at $R = 0.5$ when the design is not reliable over the interval. Data are from Supplementary Table 1.

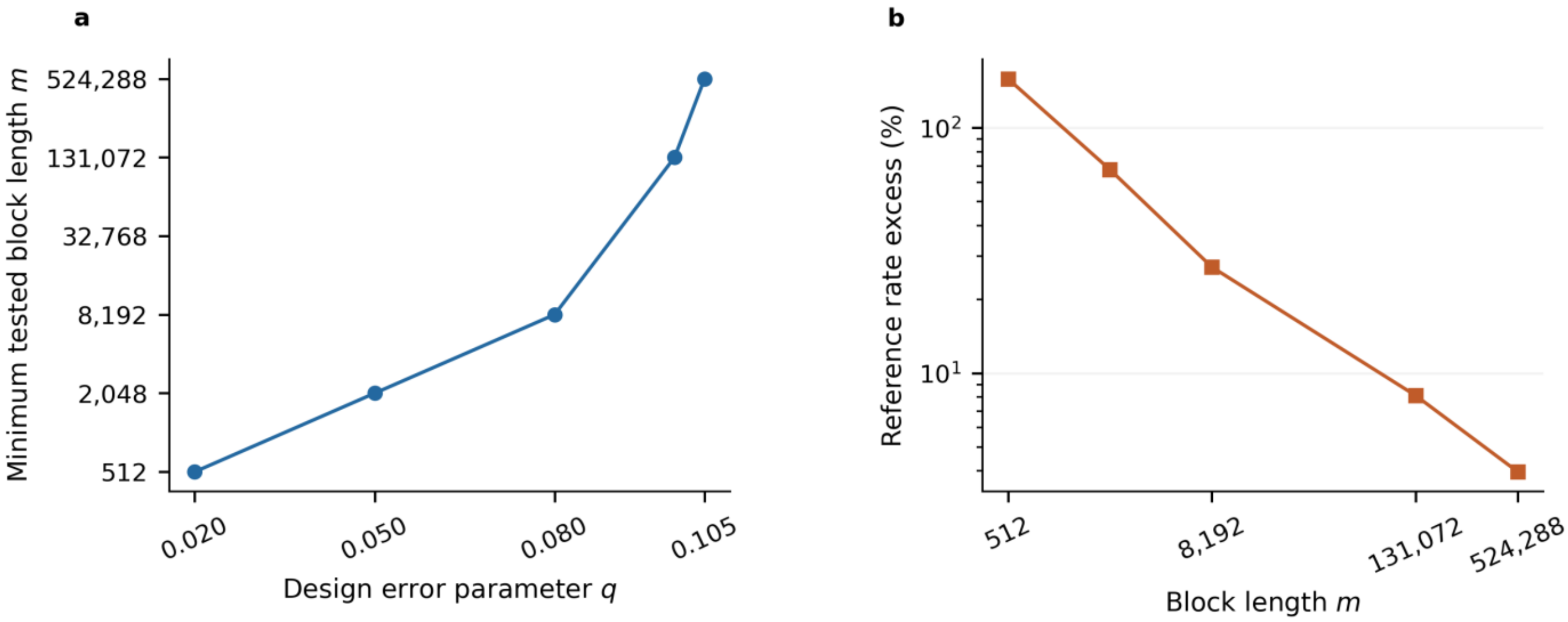

**Extended Data Fig. 2 |** Block length required to approach the dissipation frontier. a, First block length in the tested powers of two for which the whole-message failure upper bound is below $10^{-6}$, at $M = 2^{20}$, $R = 0.5$, $\rho = 100$ and $\lambda\tau = M$, for five design parameters $q$. b, Excess of the independent-bit reference dissipation rate over the asymptotic frontier, plotted for these same choices. Lines guide the eye. These are numerical evaluations of existence guarantees, not simulations of large individual codes; finite-code rate deviations are bounded in Supplementary Table 2. The minimum tested length need not be the smallest possible code length.

# Supplementary Information

## Thermodynamic limits of information storage under persistent noise

These notes supplement the model, capacity and minimum-dissipation theorems, the numerical results and the computational application in the main article. Equation numbers with an S prefix are local to this document. Unprefixed equation, figure and reference numbers refer to the main article. All entropies are in nats; physical entropy production is divided by $k_{\mathrm{B}}$. The symbol $\tau$ denotes a storage interval or a computational deadline as appropriate.

### Supplementary Note 1 | Scope of the noise assumption

#### Finite local energy and exact physical configurations

Consider an equilibrium memory with $M$ binary variables and additional degrees of freedom $A$, governed by an energy function $H(S, A)$. Suppose each bit flip is an allowed change of configuration at fixed $A$, and the absolute energy change is bounded by $E_{\max}$, independently of $M$. Here $H$ denotes energy rather than entropy, and $T$ denotes absolute temperature. For a prescribed target bit string $s^*$, the Gibbs conditional distribution satisfies

$$\Pr\big(S_i = s_i^* \mid S_{\text{-}i}, A\big) \le 1 - q_E, \qquad q_E = \frac{1}{1 + \exp(E_{\max}/k_{\mathrm{B}}T)}. \qquad \text{(S1)}$$

Conditioning successively on the preceding target bits and averaging over the remaining variables yields $\Pr\big(S = s^*\big) \le (1 - q_E)^M$, without an independence assumption. Let $\varepsilon$ denote the maximum permitted probability of incorrect readout. Requiring this probability to be at least $1 - \varepsilon$ implies

$$\frac{E_{\max}}{k_{\mathrm{B}}T} \ge \ln\frac{(1-\varepsilon)^{1/M}}{1-(1-\varepsilon)^{1/M}} = \ln M - \ln[-\ln(1-\varepsilon)] + o(1). \qquad \text{(S2)}$$

Thus exact microscopic reliability has a logarithmic protection requirement. The bound is a consequence of the finite-energy property of the Gibbs distribution, not a bound on decoded logical reliability. The bound is inapplicable to a hard-constrained state space where a single bit flip may be forbidden. The bound also does not establish any lower bound on the activation barrier or the kinetics: a finite state-energy difference need not imply a nonzero size-independent flip rate.

#### Passive versus active memories

The notation $D_{\mathrm{KL}}$ denotes Kullback–Leibler relative entropy, with initial and final distributions $p_0$ and $p_\tau$. For a fixed reversible generator with stationary distribution $\pi$, the entropy production over an interval is the decrease in relative entropy,

$$\Sigma_Z = D_{\mathrm{KL}}(p_0\|\pi) - D_{\mathrm{KL}}(p_\tau\|\pi). \qquad \text{(S3)}$$

This follows by inserting detailed balance into the channel-resolved entropy balance. A metastable memory can therefore preserve a logical message for a very long time without a maintenance cost proportional to time. For example, if the only physical variables are $M$ bits and $p_0$ is concentrated on a minimum-energy configuration of a Gibbs model, then $D_{\mathrm{KL}}(p_0\|\pi) \le M\ln 2$. Self-correcting code Hamiltonians (main article, Ref. 47) illustrate why the capacity and dissipation frontier in the main article are not universal constraints on passive memories. Our model requires a separately present symmetric noise channel on every bit even in the presence of repair. The aggregate rates of a coded active memory can obey detailed balance, while the underlying channels exchange opposite probability currents and dissipate. Equation (S3) applies to the physical channels only when those channels themselves satisfy detailed balance with the same stationary distribution. The decrease of relative entropy is also discussed in the main article, Refs. 6 and 49.

## Supplementary Note 2 | Exact and blockwise logical readout

### A transient comparison for exact physical reliability

Use the main model with an unavoidable symmetric flip rate $\lambda$ and aggregate additional flip rate at most $r$ for each bit. Fix a target string and write $E_i = 1$ for an incorrect bit. The generator can be represented by two independent Poisson clocks of rate $\lambda$ per bit, setting $E_i$ to zero or one, respectively, and a rate-$r$ repair proposal clock with state-dependent acceptance. A setting operation that leaves the bit unchanged is a null event. The representation remains valid when the repair acceptance and auxiliary state transitions depend on the full state and time. The index $i$ labels the physical bit, and $E_i$ is the associated binary error variable.

Couple this process to independent reference bits, using the same noise and proposal clocks, but set the reference bit to zero at every repair proposal. If the reference bits start at zero, they are coordinatewise no larger than the actual error bits at all later times. The reference error probability is

$$q_t = q_*\left[1 - e^{-(2\lambda+r)t}\right], \qquad q_* = \frac{\lambda}{2\lambda + r}. \qquad \text{(S4)}$$

Therefore, for any initial state and a fixed target, $\Pr\left(S_t = s^*\right) \le (1 - q_t)^M$. At long times the requirement $\Pr\left(S_t = s^*\right) \ge 1 - \varepsilon$ imposes

$$\frac{r}{\lambda} \ge \frac{1}{1 - (1-\varepsilon)^{1/M}} - 2 = \frac{M}{-\ln(1-\varepsilon)} + O(1). \qquad \text{(S5)}$$

This kinetic constraint differs from (S2). A size-independent thermal error rate and bounded repair rate are incompatible with high simultaneous physical-bit accuracy, even under arbitrary state-dependent repair. If the noise rate is Arrhenius with a fixed prefactor and $r$ is bounded, (S5) requires a barrier growing at least logarithmically with $M$; that additional Arrhenius assumption is not used in the main theorem.

For a first-passage variant, start with all bits wrong and couple the reference process from the same state. Its independent error probability is $q_* + (1 - q_*)e^{-(2\lambda+r)t}$. If $Mq_* \ge 1$, the incoming probability flux into the all-correct state is at most $(\lambda + r)Mq_*(1 - q_*)^{M-1}$, because $Mq(1-q)^{M-1}$ is decreasing for $q \ge 1/M$. Let $T_{\text{hit}}$ be the actual first hitting time of the target. The coupling and the expected number of entries into the reference target give

$$\Pr(T_{\text{hit}} \le t) \le Jt, \qquad J = (\lambda + r)Mq_*(1 - q_*)^{M-1}. \qquad \text{(S6)}$$

Integrating the resulting survival bound on $[0,1/J]$ gives $\mathbb{E}T_{\text{hit}} \ge 1/(2J)$. At fixed positive $\lambda, r$ this grows at least exponentially, up to a polynomial factor, with $M$. This is a bound for exact physical configurations, not for arbitrary decoded memories, and does not assume the ability to detect the target for free in a real device.

### Disjoint block decoders

Suppose each of $K$ logical bits is decoded from a disjoint block of $m_i$ physical bits ($\sum_i m_i = M$), and each block decoder can output both logical values. In the Gibbs model above, every specified pattern of $m_i$ bits has conditional probability at least $q_E^{m_i}$ given the bits outside its block. Choose a pattern decoded incorrectly in each block. Sequential conditioning yields an upper bound $\prod_i \left(1 - q_E^{m_i}\right)$ on full logical success. In the dynamical model, an analogous bound holds at time $t$: the event that the most recent setting or repair proposal for a given bit was the noise setting to a prescribed value has probability $q_t$. These events are independent across distinct bits, even though the resulting physical states need not be. Consequently the event that all bits in a block were last set to the selected wrong pattern has probability $q_t^{m_i}$. The construction supplies independent sufficient events for a logical error in the distinct blocks. Both bounds have the form

$$\Pr(\text{all logical bits correct}) \leq \exp\left[-Kq_0^{M/K}\right], \qquad \text{(S7)}$$

where $q_0 = q_E$ or $q_t$, respectively. The exponential form follows from $1 - z \leq e^{-z}$ and convexity of $q_0^m$. Thus a necessary condition for success probability $1 - \varepsilon$ is

$$\frac{M}{K}\ln\frac{1}{q_0} \geq \ln\frac{K}{-\ln(1-\varepsilon)}. \qquad \text{(S8)}$$

At a fixed nonzero $q_0$ this requires $M = \Omega(K\ln K)$. The order is attainable for independent Bernoulli errors with a fixed probability $q < 1/2$ using odd-length repetition blocks: the majority error of a block of length $m$ is bounded by $\exp\left[-mD_{\mathrm{KL}}\left(\mathrm{Bern}(1/2)\|\mathrm{Bern}(q)\right)\right]$, and a union bound suffices. The construction concerns readout of already encoded information, not the physical cost of a fault-tolerant encoder or logic gates. In contrast, the constant-rate joint codes in the main article protect many logical bits in one block. The restriction to disjoint blocks each decoding only one logical bit is essential to (S8); the bound does not apply to the block codes in the main article, each of which encodes a number of logical bits proportional to its length. Supplementary Fig. 1 illustrates the distinct scaling of exact physical readout and disjoint one-bit repetition blocks. Standard coding and binomial-tail bounds are described in the main article, Ref. 35.

### Maintenance entropy production for a prescribed configuration

For a fixed target physical string, let $P_C$ be the probability of the target and $P_1$ the probability of being at Hamming distance one. Under the noise channel, the fluxes out of and into the target set are $\lambda MP_C$ and $\lambda P_1$, respectively. The log-sum inequality gives

$$\dot{\Sigma}_{\mathrm{noise}} \geq \lambda(MP_C - P_1)\ln\frac{MP_C}{P_1}. \qquad \text{(S9)}$$

For $P_C \geq 1 - \varepsilon$ and $M(1-\varepsilon) > \varepsilon$, this is at least $\lambda[M(1-\varepsilon) - \varepsilon]\ln[M(1-\varepsilon)/\varepsilon]$. Thus exact physical retention requires an entropy production rate of order at least $\lambda M\ln M$ when attainable. The condition (S5) must also hold in the long-time limit. In an independent-bit example with repair rate $r$ increasing proportionally to $M$ and fixed reverse rate $b > 0$, the physical stationary bit error $q = (\lambda + b)/(2\lambda + r + b)$ is of order $1/M$ and the stationary rate $\lambda M(1 - 2q)\ln(r/b)$ has the same order. This example is outside the fixed-$r$ asymptotic model of the main article. When the target depends on a message label rather than being a fixed physical configuration, (S9) can instead be applied conditionally, with the entropy-balance correction in equation (14) of the main article.

## Supplementary Note 3 | Finite-time bounds and limiting regimes

The message label is not a thermodynamic resource continually accessible to the device. Conditioning on that label changes the system entropy but not the mean medium entropy production if the generator is common to all messages. Equation (14) follows from $H(Z) = H(Z \mid X) + I(X; Z)$. For example, a single bit initially equal to a uniformly random message and then allowed to undergo symmetric noise has a uniform physical marginal at every time. Its physical entropy production is zero, even though $I(X; S_t)$ decays. The example explains why the boundary term in equation (6) cannot be omitted in general at finite times.

### Simpler bounds for the rate frontier

With $d = \ln 2 - h(q)$ and $\phi(d) = (1 - 2q)\ln[(1-q)/q]$ as in Methods, the identity $d\phi/dd = 2 + 2\sinh(x)/x$ implies $\phi(d) \geq 4d$. Jensen's inequality and the conditional entropy chain rule then give the conditional noise entropy production bound

$$\dot{\Sigma}_{\text{noise},Z|X} \geq \lambda M \phi\left(\frac{I(X;S)}{M}\right) \geq 4\lambda I(X;S). \qquad \text{(S10)}$$

Using equation (14), the corresponding lower bound for the total physical entropy production is

$$\Sigma_Z \geq 4\lambda \int_0^\tau I\,(X;S_t)\,dt + I(X;Z_\tau) - I(X;Z_0). \qquad \text{(S11)}$$

This linear bound applies without specifying the encoding or a stationary distribution. The inequality concerns entropy production, not merely heat transferred to the bath. The sharper inequality (6) additionally includes the finite-rate penalty through the function $\Psi_\rho$. The instantaneous noise entropy production rate in equation (S10) is distinguished from integrated entropy production. The underlying binary entropy inequality is related to the improved log-Sobolev inequality (main article, Ref. 36).

For completeness, a weaker finite-rate bound can be obtained without the Legendre transform. Use the physical repair fluxes per conditional weight (not normalized by $\lambda$) and write $j = F - B$. For $j > 0$, $j\ln(F/B) \geq j^2/F \geq j^2/(rq)$. If $\mathcal{E}$ is the net rate at which repair decreases $H(S \mid X)$, Cauchy–Schwarz implies $\dot{\Sigma}_{\text{repair},Z|X} \geq [\mathcal{E}]_+^2/(rMC_0)$, where

$$C_0 = \max_{0<q\leq 1/2} q\left[\ln\frac{1-q}{q}\right]^2. \qquad \text{(S12)}$$

Here $[a]_+ = \max(a, 0)$. The bound follows by retaining only positive repair currents and weighting the square by $rq$. The time integral of $\mathcal{E}$ equals the conditional noise entropy increase minus $\Delta H(S \mid X)$. The constant $C_0$ is dimensionless. Combining with (S10), for a mean information per bit at least $d_0 = R_0\ln 2$,

$$\frac{\Sigma_Z}{\lambda M\tau} \geq \phi(d_0) + \frac{1}{\rho C_0}\left[\phi(d_0) - \frac{\ln 2}{\lambda\tau}\right]_+^2 - \frac{K\ln 2}{\lambda M\tau}. \qquad \text{(S13)}$$

The sharp bound (6) is used for the main results instead of (S13). In particular the weaker inequality alone does not establish the divergence at $R_*$. In the high-reliability long-time limit, at $d_0 = R\ln 2$, the right-hand side of (S13) becomes $\phi(R\ln 2) + \phi(R\ln 2)^2/(\rho C_0)$.

## Explicit slope and singular limits

The supporting function in Methods has positive slope at the point $d_0$. Differentiating the balanced-current cost explicitly, with $q_0 = q_{R_0}$ and $\ell_0 = \ln[(1-q_0)/q_0]$, gives

$$\begin{aligned} L'_\theta(d_0) &= \frac{2\ln A_0}{\ell_0} + \frac{1-2q_0}{\ell_0}\left[\frac{1}{1-q_0} + \frac{\rho+2}{(\rho+2)q_0-1}\right], \\ A_0 &= \frac{\rho(1-q_0)}{(\rho+2)q_0-1}. \end{aligned} \qquad \text{(S14)}$$

All terms are positive in the open interval $0 < R_0 < R_*$. The numerical audit in the accompanying program evaluates the tangent inequalities for the capacity function and the supporting function at six rate ratios (0.1, 1, 3, 10, 100 and 1,000), 25 contact points per ratio and 4,000 evaluation values of $q$ per contact. The contact values range from 0.05 to 0.99 times $R_*$. The evaluation combines a logarithmic and a linear grid in $q$, as specified in the program. The 600,000 tests give no negative supporting-line residual within the stated numerical tolerances; the largest absolute contact error is $1.42 \times 10^{-14}$. These checks supplement rather than replace the analytical proof in Methods.

As $R \downarrow 0$ at fixed $\rho$, $d = x^2/8 + O(x^4)$ and $\Psi_\rho(R)/R \to 4\ln 2(1 + 1/\rho)$. As $R \uparrow R_*$, $q_R - q_* \sim (R_* - R)\ln 2/\ln[(1 - q_*)/q_*]$. The reverse repair flux of the attaining process tends to zero, while the net current tends to the nonzero value $\lambda(1 - 2q_*)$. This yields the logarithmic singularity in equation (5). At a fixed $R_0 < R_*$ the boundary terms in equation (6) are uniformly $O(M)$, even if the number of auxiliary states grows. Taking the long-time limit at fixed $R_0$ before taking $R_0 \uparrow R_*$ avoids a nonuniform limit of the supporting coefficient $\theta_{R_0}$.

## Supplementary Note 4 | Syndrome-space calculation

Let $H$ be a full-rank parity-check matrix with binary column vectors $h_i$. A physical flip of bit $i$ maps the syndrome $y$ to $y \oplus h_i$. Let $e_y$ be the chosen minimum-weight coset leader, obtained by breadth-first search from syndrome zero, and let $w_y = |e_y|$. The stationary syndrome probability is proportional to $[q/(1-q)]^{w_y}$, since each syndrome has exactly $2^K$ representatives in the physical space. Its transition rate for bit $i$ is $\lambda + r$ if $w_{y \oplus h_i} < w_y$, and $\lambda + b$ otherwise. Distinct bits are retained as distinct transitions when computing entropy production. The notation wt denotes Hamming weight; here w(y) is the weight of the coset leader.

The decoded message changes if and only if the $M$-bit vector $e_y \oplus e_{y \oplus h_i} \oplus u_i$ is nonzero, where $u_i$ is the unit vector for bit $i$. This vector belongs to the code. Denote this indicator by $J_{yi}$, the corresponding aggregate rate by $w_{yi}$, and the normalized stationary syndrome probabilities by $\pi_y$. The decoder-change hazard in equation (11) can be calculated directly as

$$\kappa = \sum_y \pi_y \sum_i w_{yi} J_{yi}. \qquad \text{(S15)}$$

The instantaneous decoding success for a binary symmetric channel (BSC) with design error $q$ is $\sum_y q^{w_y}(1 - q)^{M\text{-}w_y}$. The channel-decoding success is not the probability of correct decoding at a random time relative to the original message in the long-time stationary physical ensemble. In the latter ensemble the initial message is defined as the decoded value at the start of storage. The retention-failure event is any change of that decoded value before the deadline.

For exact first-passage evaluation, construct a column generator $Q$ with off-diagonal rates only for transitions with $J_{yi} = 0$, and keep the full escape rate in the diagonal (including transitions with $J_{yi} = 1$). If all columns are distinct nonzero parity-check vectors, no self-flips of the syndrome occur, as in our sampled matrices. The resulting killed process has survival probability

$$P_{\text{survive}}(\tau) = \mathbf{1}^{\mathrm{T}} \exp(Q\tau)\boldsymbol{\pi}. \qquad \text{(S16)}$$

Supplementary Table 1 summarizes the numerical results. For larger matrices, uniformization uses a Poisson number of proposals with mean $M(\lambda + r)\tau$, each choosing a uniformly random bit and accepting with probability $w_{yi}/(\lambda + r)$. A trajectory is killed on the first accepted proposal with $J_{yi} = 1$. Uniformization generates the continuous-time process exactly, up to random sampling error. The code verifies the coset leaders, probability normalization and aggregate detailed balance. The random matrix and trajectory seeds are specified in Methods; the complete code is supplied separately.

**Supplementary Table 1 | Additional finite-size results.** The density is $R = 0.5$ and $\lambda\tau = 1$. The BSC error is evaluated by summing all minimum-weight leaders; the failure probability concerns any decoder change during the interval. The entropy production rate is normalized by $\lambda M$. E denotes exact killed-generator evaluation and MC denotes Monte Carlo evaluation with 5,000 uniformization trajectories. Values of one in the failure column are rounded or sample estimates, not proofs of certain failure. For MC rows the standard errors are binomial estimates; when all trajectories fail these estimates are zero but do not imply zero uncertainty. These are individual fixed code realizations, not ensemble averages.

| $M$ | $\rho$ | $q$ | BSC error | Failure | MC standard error | $\sigma/(\lambda M)$ | Method |
|---|---|---|---|---|---|---|---|
| 16 | 10 | 0.090 | 0.202746 | 0.989622 | – | 3.629219 | E |
| 16 | 100 | 0.020 | 0.011589 | 0.943550 | – | 4.338978 | E |
| 16 | 100 | 0.100 | 0.242185 | 1.000000 | – | 1.687846 | E |
| 24 | 10 | 0.090 | 0.212311 | 0.992603 | – | 3.455748 | E |
| 24 | 100 | 0.020 | 0.007055 | 0.840689 | – | 4.299645 | E |
| 24 | 100 | 0.100 | 0.261484 | 1.000000 | – | 1.597647 | E |
| 32 | 10 | 0.090 | 0.198333 | 0.991200 | 0.001321 | 3.611521 | MC |
| 32 | 100 | 0.020 | 0.002124 | 0.560000 | 0.007020 | 4.356362 | MC |
| 32 | 100 | 0.100 | 0.255417 | 1.000000 | 0.000000 | 1.671716 | MC |
| 40 | 10 | 0.090 | 0.188008 | 0.992800 | 0.001196 | 3.613862 | MC |
| 40 | 100 | 0.020 | 0.000786 | 0.305400 | 0.006514 | 4.359022 | MC |
| 40 | 100 | 0.100 | 0.251420 | 1.000000 | 0.000000 | 1.670290 | MC |

At $\rho = 100$ and $q = 0.1$, several small memories dissipate less than the asymptotic frontier at $R = 0.5$. The observation is not a violation of the bound: they do not reliably retain the initial message over the interval. The finite-time result is expressed in terms of mutual information retained over the interval and includes its boundary correction, not the nominal code rate alone.

## Supplementary Note 5 | Block-size and dissipation guarantees

For a block of length $m$ and dimension $k$, equation (10) gives the ensemble bound $\delta_m$ on the Bernoulli bad-set mass. The same code satisfies the finite-size stationary dissipation guarantee in Methods. For $J = M/m$ independent blocks and $K = Jk$, add their entropy production rates and apply a union bound over their first decoder changes. In particular the normalized whole-memory dissipation differs from the independent-bit reference value by at most

$$\left(2\delta_m + \frac{2\delta_m}{1-\delta_m}\right)\ln\frac{r}{b}. \qquad \text{(S17)}$$

Supplementary Table 2 reports the block-size bounds. The same $\delta_m$ also bounds the failure probability through equation (11) with $M$ replacing the block length in the prefactor. The estimates are useful near the frontier because they control both reliability and dissipation for one and the same code; using separate ensemble-optimal codes for the two quantities would not establish joint attainability.

**Supplementary Table 2 | Block construction at fixed total size.** $M = 2^{20}, K = M/2, \rho = 100$ and $\lambda\tau = M$. The block size is the first power of two in the tested range $128, 256, \ldots, M$ for which the upper bound on the whole-message failure probability is less than $10^{-6}$. The normalized rate shown is the independent-bit reference value $\psi(q) =$

$(1-2q)\ln(r/b)$, and the final column is an upper bound on its absolute discrepancy from the finite-code stationary rate. Excess is relative to $\Psi_{100}(0.5) = 1.68778413$. All entries are numerical evaluations of existence bounds, not simulations of individual large codes.

| $q$ | $m$ | Blocks | Failure bound | $\psi(q)$ | Excess (%) | Rate error bound |
|---|---|---|---|---|---|---|
| 0.020 | 512 | 2,048 | $1.57 \times 10^{-9}$ | 4.363917 | 158.56 | $2.56 \times 10^{-22}$ |
| 0.050 | 2,048 | 512 | $2.73 \times 10^{-13}$ | 2.828601 | 67.59 | $3.09 \times 10^{-26}$ |
| 0.080 | 8,192 | 128 | $1.07 \times 10^{-7}$ | 2.144754 | 27.08 | $9.83 \times 10^{-21}$ |
| 0.100 | 131,072 | 8 | $4.48 \times 10^{-19}$ | 1.824485 | 8.10 | $3.68 \times 10^{-32}$ |
| 0.105 | 524,288 | 2 | $2.26 \times 10^{-18}$ | 1.754655 | 3.96 | $1.80 \times 10^{-31}$ |

The bounds are evaluated using logarithmic binomial coefficients and log-sum-exp, as in Methods. The rate error bounds are analytically derived estimates evaluated numerically; they are not claims of floating-point resolution of the rate to that many digits. The required block size grows sharply as $q$ approaches $q_R$. The logarithmic asymptotic block-size scaling holds at fixed positive distance from the frontier and does not guarantee a practical small block length or a fixed-degree physical repair network.

## Supplementary Note 6 | Computation and information retention

### Transport bound and a passive-computation comparison

Let $p_t$ be a distribution on a graph of progress states, with integrated expected jump count $\mathcal{A}$. At each time, write forward and backward fluxes of an undirected edge as $a, b$. Its entropy production is $2|a-b|\operatorname{artanh}(|a-b|/(a+b))$. Convexity of the function $2x\operatorname{artanh}x$ on $0 \le x < 1$ gives, after averaging over edges and time with activity weights, a bound $2J\operatorname{artanh}(J/\mathcal{A})$, where $J$ is the time integral of the sum of absolute net edge currents. The continuity equation implies $J \ge W_1(p_0, p_\tau)$ for unit graph edge lengths (main article, Ref. 40). Thus

$$\Sigma_{\text{prog}} \ge 2W_1\operatorname{artanh}(W_1/\mathcal{A}) \ge \frac{2W_1^2}{\mathcal{A}}. \qquad \text{(S18)}$$

For a deterministic start and final success probability $1-\varepsilon_c$ in a set at graph distance at least $L$, $W_1 \ge L(1-\varepsilon_c) = D$. If the aggregate progress-changing rate is capped by $r_c$, then $\mathcal{A} \le r_c\tau$. Equation (12) follows by adding the bound on disjoint memory channels from equation (6), with the message-ensemble boundary correction counted once. The sum is not justified if the progress transitions also update the noisy register $S$.

In a purely passive reversible implementation with fixed stationary $\pi$, equation (S3) also caps the available entropy production by $D_0 = D_{\text{KL}}(p_0\|\pi)$. For the same distance and activity constraints, (S18) then implies

$$\tau \ge \frac{2L^2(1-\varepsilon_c)^2}{r_c D_0}. \qquad \text{(S19)}$$

The bound gives a diffusive lower-bound scaling when $L$ increases linearly with problem size and $D_0$ and $r_c$ remain bounded. It is not a general statement that equilibrium computation is diffusive; a growing initial free-energy resource, parallel transitions, or a different state graph can change the scaling. Non-equilibrium currents can add entropy production beyond the relaxation budget. The clocked example in the main article demonstrates the transport term with logical operations, subject to the explicitly ideal accumulator and gate assumptions.

## Information-time volume and the release of unnecessary memory

Equation (S11) penalizes the integral of retained information rather than only the number of logical gates. Consider a sequential process of duration proportional to $L$ that retains $K = \Theta(L)$ input bits reliably throughout, with $M = \Theta(K)$. For fixed nonzero noise and repair rates, its maintenance contribution is $\Omega(L^2)$. The conclusion uses the channel separation and information-retention assumptions, not logical reversibility alone. If inputs are reliably streamed from outside the device and only a bounded amount of information must be retained internally, the internal information-time lower bound can instead scale linearly. Any external storage, reliable data supply and output-protection costs must then be accounted for separately.

A simple release comparison can be made without claiming that irreversible logic is necessarily more efficient. Take the stationary coded memory in the main article, with physical distribution $\pi$. Once its message is no longer needed, turn off the repair while leaving the symmetric noise active. The memory relaxes to the uniform distribution and produces at most $M\ln 2 - H(\pi)$ in entropy production, independent of the duration. If the same physical memory is instead maintained in $\pi$ for an additional duration $t_h$ and then released, its maintenance contribution is $\sigma_{\mathrm{ss}} t_h$ in addition to the same eventual relaxation cost. Thus delaying the release of unneeded information in this example has a real thermodynamic cost. The comparison is made at the same eventual equilibrium state; it does not include the cost of resetting to a prescribed fresh codeword, changing the control mechanism, or the finite deadline required for thermalization. Neither the release nor its cost should be identified with Landauer erasure to a specified logical state.

## Numerical checks of the clock contribution

Use the reflecting clock of Methods with $r_c = 1$, $\delta = 0.5$, $\delta_L = \delta + L^{-1/3}$ and $\tau = L/\delta$. The two reflecting endpoints are $-L$ and $2L$. The tridiagonal generator propagates the distribution from $h = 0$. The exact marginal clock entropy production is $\langle h_\tau \rangle \ln(u/v) + H(h_\tau)$, which excludes the stationary memory maintenance and the at most ln2 system-entropy contribution from the ideal accumulator. We compare it to the lower bound with $D = L\Pr(h_\tau \geq L)$ and to the limiting drift cost $2L\operatorname{artanh}\delta$. The normalized probability and column sums of the generator are checked in the accompanying program. The completion probability and the convergence of the clock cost illustrate the analytic limit, not the performance of a physical gate or decoder. Supplementary Fig. 2 visualizes the finite-clock convergence in Supplementary Table 3.

**Supplementary Table 3 | Finite-clock dynamics.** Time is in units of $1/r_c$ and entropy production in units of $k_{\mathrm{B}}$. $\Sigma_{\mathrm{clock}}$ is the marginal clock entropy production; $\Sigma_{\mathrm{lower}}$ is equation (S18) with the deadline success distance and the rate cap. The last column is the ratio to the limiting leading-order drift cost. The finite-size drift increment $L^{-1/3}$ deliberately makes completion reliable, at the expense of excess dissipation at finite $L$.

| $L$ | $r_c\tau$ | Completion | Mean position | $\Sigma_{\mathrm{clock}}$ | $\Sigma_{\mathrm{lower}}$ | Ratio |
|---|---|---|---|---|---|---|
| 64 | 128 | 0.998452 | 95.9891 | 190.623 | 70.0705 | 2.7111 |
| 256 | 512 | 0.999866 | 336.635 | 535.330 | 281.161 | 1.9034 |
| 1,024 | 2,048 | 0.999997 | 1,227.187 | 1,703.457 | 1,124.972 | 1.5142 |

No inference of a universal optimal computing architecture follows from this example. Even when the normalized memory maintenance cost approaches the frontier, a subleading difference per bit can accumulate over a large memory and a long interval. The example establishes leading-order attainability of the transport contribution relative to the *actual* stationary maintenance cost, not necessarily a second-order expansion about the optimum of the joint cost.

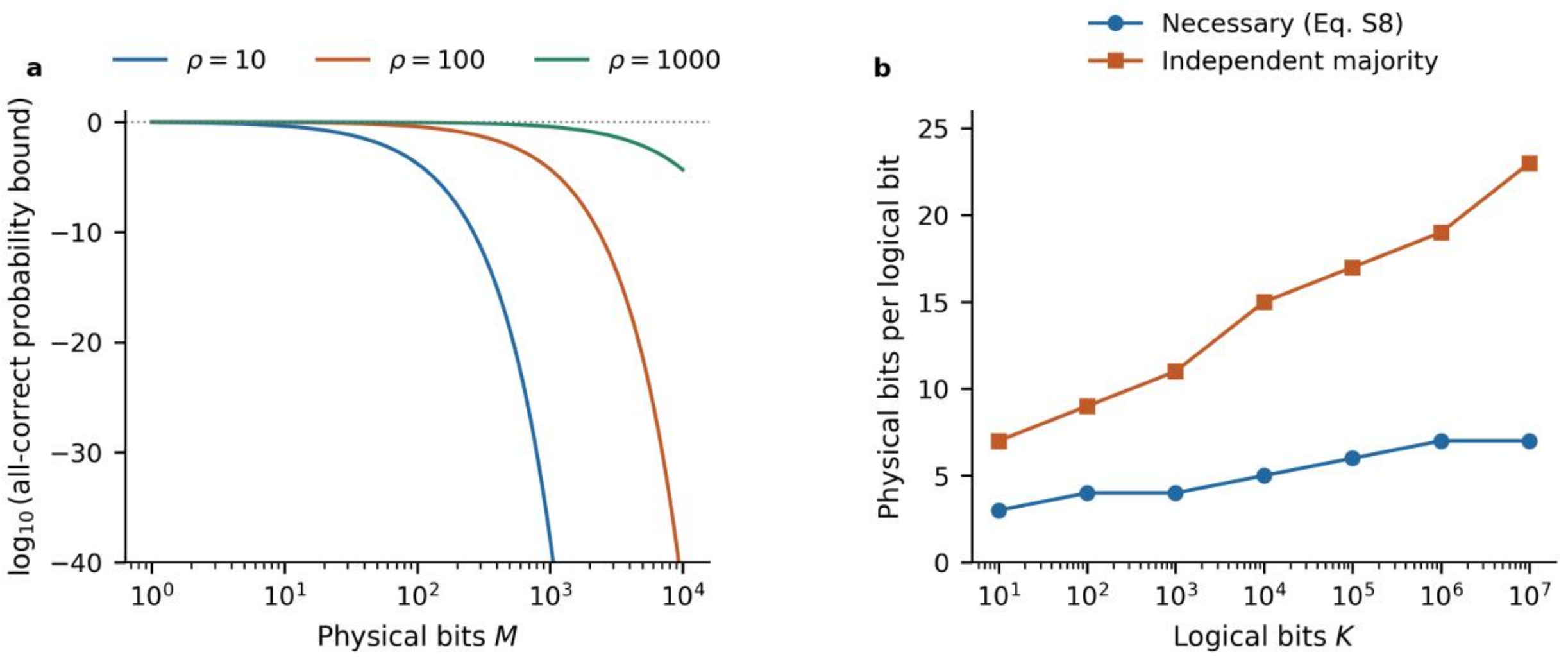

**Supplementary Fig. 1 |** Exact physical readout versus disjoint logical decoding. a, Long-time upper bound on the probability that all physical bits match a prescribed target, from equation (S4), at three repair-to-noise ratios. Values below the plotted range are not shown. Dotted line: 99% target success. b, Necessary mean block length (integer ceiling of equation (S8)) and minimum odd repetition length calculated from independent binomial errors, for $q_0 = q = 0.05$ and 99% joint logical success. The latter is a readout construction for independent errors; it does not model the physical gate implementation. These curves illustrate Supplementary Note 2.

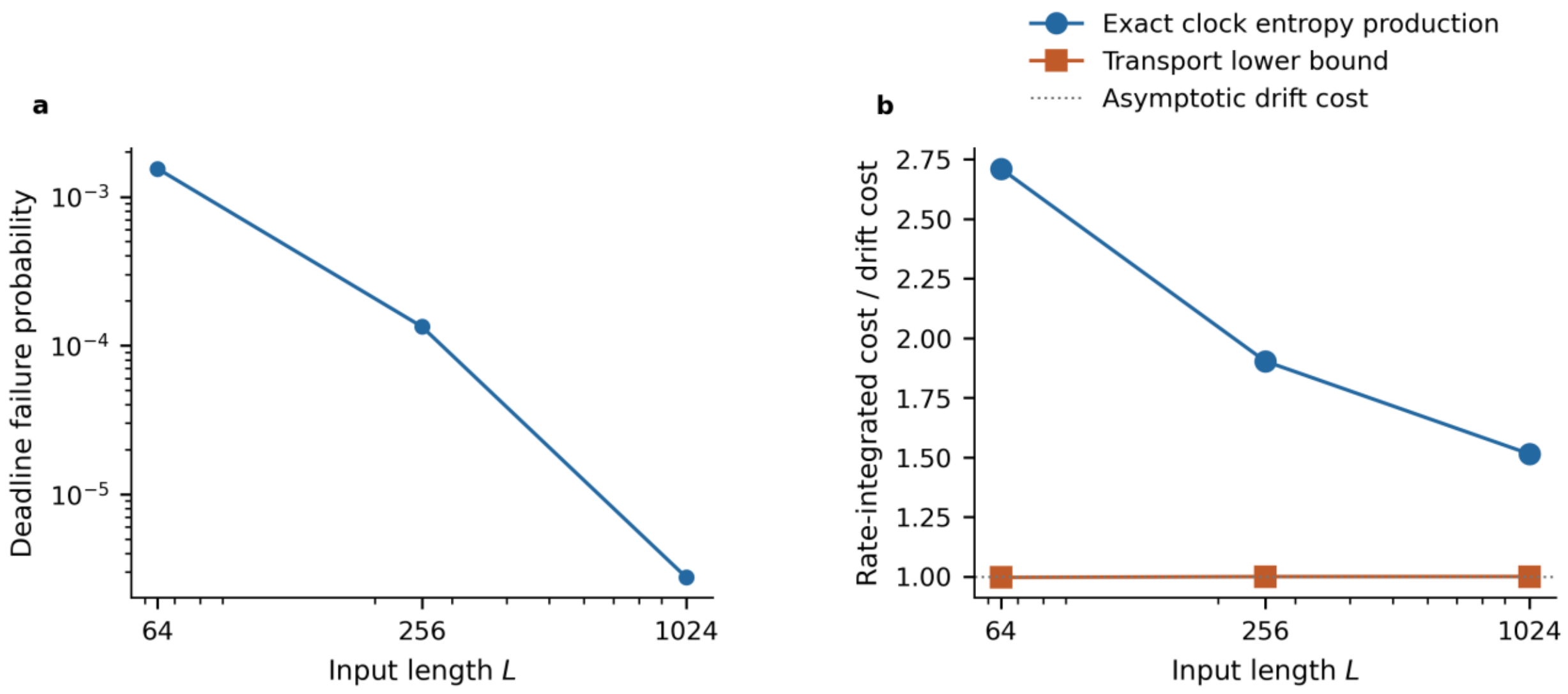

**Supplementary Fig. 2 |** Numerical illustration of the transport contribution. a, Deadline failure probability of the finite reflecting clock in Supplementary Table 3. b, Exact marginal clock entropy production and its transport lower bound, normalized by the leading drift cost $2L$ artanh(0.5). The dotted line is the asymptotic prediction. Values are computed for the finite-size drift increment described in Supplementary Note 6; they exclude memory maintenance and ideal gate or accumulator costs.